\documentclass[final,3p,times]{elsarticle}

\usepackage{amssymb}
\usepackage{amsmath}

\usepackage{multirow}%
\usepackage{amsmath,amssymb,amsfonts}%
\usepackage{amsthm}%
\usepackage{mathrsfs}%
\usepackage{textcomp}%
\usepackage{manyfoot}%
\usepackage{booktabs}%
\usepackage{algorithm}%
\usepackage{algorithmicx}%
\usepackage{algpseudocode}%
\usepackage{listings}%
\usepackage[T1]{fontenc}
\usepackage{import}

\usepackage{subcaption}
\usepackage{layouts}
\usepackage{lmodern}
\usepackage{siunitx}    
\DeclareSIUnit{\wtpercent}{wt\%}
\DeclareSIUnit{\voxel}{vx}
\DeclareSIUnit{\px}{px}
\usepackage[size=scriptsize]{todonotes}
\usepackage{wrapfig}
\usepackage{pgfplots}
\pgfplotsset{compat=1.18}
\usepgfplotslibrary{groupplots}
\usepackage{import}
\usepackage{multirow}
\usepackage{rotating}

\usepackage{pifont}
\usepackage{tabularx}
\usepackage{natbib}
\usepackage{booktabs}
\usepackage{float}
\journal{Experimental Thermal and Fluid Science}

\begin{document}

\begin{frontmatter}

\title{Influence of the Inhalation Route on Tracheal Flow Structures in Patient-Specific Airways using 3D PTV}

\author[AIA]{Benedikt H. Johanning-Meiners}
\affiliation[AIA]{organization={Institute of Aerodynamics and Chair of Fluid Mechanics, RWTH Aachen University},
            addressline={W{\"u}llnerstr.~5a},
            city={Aachen},
            postcode={52062},
            state={North~Rhine-Westphalia},
            country={Germany}}
\author[AIA]{Luca Mayolle}
\author[AIA]{Dominik Krug}
\author[AIA]{Michael Klaas}

\begin{abstract}
The tracheal flow field shapes particle transport into the lower airways and thus influences both the spread of inhaled pathogens and the effectiveness of aerosol-based therapies. Identifying how different inhalation routes modify the flow field is therefore crucial for understanding lower-airway disease transmission and for guiding targeted drug delivery. To gain a detailed understanding of the influence of the inhalation route on the flow structures in the human trachea, the three-dimensional flow field in the trachea is investigated \textit{in vitro} in a non-compliant, refractive-index matched silicone model of the human respiratory tract. The investigations comprise steady inflow conditions, i.e., steady inhalation, and oscillatory flow combining inhalation and exhalation to simulate calm breathing. A realistic breathing pattern is approximated by a sinusoidal waveform for two Reynolds numbers of $Re_{Tr} = [400, 1200]$, based on the bulk velocity at maximum volume flux and the hydraulic diameter of the trachea and two Womersley numbers of $Wo = [3, 4.5]$, representing the oscillation time scales. To capture the inherently three-dimensional and asymmetric nature of the flow field, 3D particle-tracking velocimetry (PTV) measurements are performed using the Shake-The-Box (STB) algorithm. Using a refractive-index matched fluid consisting of water and glycerin, the complex flow structures inside the trachea are fully resolved. The PTV measurements confirm that the nasal and/or oral cavity must be considered when analyzing the flow field in the lower respiratory tract. In particular, we find that the presence of both cavities significantly alters the flow field compared to idealised, fully developed inflow conditions. However, velocity profiles in the sagittal and coronal plane in the trachea as well as contour plots of the of the normalized velocity magnitude evidence nearly identical flow structures for oral and nasal inhalation, indicating minimal influence of the inhalation route.   
\end{abstract}



\end{frontmatter}



\section{Introduction}
\label{Sec:Intro}
Recent years showed a significant increase in deaths from widespread respiratory diseases, such as chronic lower respiratory diseases, pneumonia, and influenza, with these diseases being responsible for approximately \SI{7.2}{\percent} of deaths in 2019~\citep{Xu.2021}. The COVID-19 pandemic has since highlighted even more sharply how respiratory pathogens can spread through inhaled aerosols, drawing renewed attention to the mechanisms by which particles are transported and deposited within the airways. SARS-CoV-2 has demonstrated that small aerosol particles can remain suspended, travel through indoor air, and reach deeper regions of the respiratory tract, where the transmission and infection is strongly governed by the underlying flow structures. While airborne transmission is generally not considered the most efficient route of contagion~\citep{Wang.2021}, it nevertheless represents a critical pathway for numerous clinically significant pathogens, including Respiratory Syncytial Virus (RSV)~\citep{Kulkarni.2016}, varicella (chickenpox)~\citep{Tang.2005}, and variola (smallpox)~\citep{Milton.2012}. Consequently, a detailed analysis of airflow dynamics and mass transport within the respiratory tract is a prerequisite for understanding the spreading mechanisms of respiratory infections. In this context, the intricate and highly three-dimensional geometry of the airways possesses a distinct influence on particle and aerosol transport and deposition in the airways, emphasizing the need for a thorough investigation of flow structures in these regions. Usually, earlier experimental and numerical investigations of respiratory airflow simplified the geometry of the conducting airways and omitted the upper regions, namely the nasal or oral cavity, pharynx, and larynx and focused on the trachea and the initial generations of the bronchial tree. As a consequence, the inflow had to be approximated typically by assuming a fully developed and laminar velocity profile at the tracheal inlet \citep{Soodt.2012, Koullapis.2018}. 

Banko et al.~\cite{Banko.2015} investigated steady inspiratory flow in a patient-specific model of the human airways extending from the mouth to the eighth bifurcation generation. The transparent model, derived from computerized tomography (CT) data of a healthy adult, was studied using magnetic resonance velocimetry (MRV) to obtain three-dimensional, three-component mean velocity fields at a Reynolds number of $Re=4200$. The measurements indicate that the geometry, i.e., the larynx and the glottis, generates a strong, asymmetric jet and a single-sided swirl that persists throughout the trachea and into the main bronchi. The swirl, combined with the inclined tracheal orientation and asymmetric bifurcation geometry, causes highly skewed velocity profiles and complex secondary vortices. The results demonstrate that upper airway morphology plays a central role in shaping tracheal flow structures and secondary motion, and therefore should not be oversimplified through idealized inlet conditions. Furthermore, these flow features interact closely with the oscillatory inhalation, highlighting that the assumption of steady inflow conditions cannot capture the physiological dynamics of a breathing cycle. R\"uttgers et al.~\cite{ruettgers.2025} conducted direct numerical simulations (DNS) of inspiratory airflow through a realistic, patient-specific airway model extending from a nasal mask down to the sixth bronchial bifurcation. The simulations were conducted at two physiologically relevant Reynolds numbers ($Re = 400$ and $Re = 1200$) using a lattice--Boltzmann method (LBM) implemented in the \textit{m-AIA} framework. The study revealed that the upper airway geometry, particularly the nasal cavity, pharynx, and larynx, crucially determines the flow entering the trachea, i.e., narrow nasal passages and curved turbinates that produce strong local accelerations and secondary vortices, while the nasopharyngeal bend and glottal region induced shear-layer instabilities and a jet-like flow (glottal jet). The authors conclude that a strong spatial correlation exists between pressure loss and the onset of flow instabilities, showing that the upper airway morphology directly governs the organization of the tracheal flow field. 

Complementing these findings, Lizal~et~al.~\cite{Lizal.2020} investigated the effect of oral and nasal breathing on the deposition of inhaled particles using a patient-specific airway replica manufactured via stereolithography and selective laser sintering based on CT scans of a healthy individual. The model was designed for flow measurements using Laser-Doppler Anemometry (LDA) and particle deposition measurements using Positron Emission Tomography (PET). The experiments were conducted for steady oral, nasal, and combined breathing routes at flow rates ranging from \SI{15}{\liter\per\minute} to \SI{60}{\liter\per\minute}. The study aimed to provide experimental data to validate numerical simulations and the experimental results confirmed that a high pressure drop across the nasal cavity leads to nearly identical velocity profiles downstream of the glottis for both oral and combined breathing. The distribution of deposited particles downstream of the trachea is unaffected by the inhalation route. Xu~et~al.~\cite{Xu.2020} investigated a 2:1 sized model of the upper airways that was reconstructed from magnetic resonance imaging (MRI) of a healthy adult and fabricated by stereolithography using PIV. While the nasal, oral, pharyngeal and laryngeal segments were made from photosensitive resin, the trachea was replaced by a high-borosilicate glass model to ensure high optical transmission. Steady air flows equivalent to \SI{18}{\liter\per\minute}, \SI{32}{\liter\per\minute}, and \SI{45}{\liter\per\minute} in vivo (or \SI{36}{\liter\per\minute}, \SI{64}{\liter\per\minute}, and \SI{90}{\liter\per\minute} in the model) were driven through oral, nasal, and combined inlets. The oral and nasal inhalation produce an anterior glottal jet and separation, and overall, the two-dimensional measurements of the flow field in the upper trachea shows similar results for both oral and nasal inhalation. By increasing the Reynolds number and thus, the flow rate, the jet phenomenon is reduced which leads to a smaller recirculation zone. 
Janke~et~al.~\citep{Janke.2017} investigated the oscillatory flow field in a generic human airway model based on the geometries of Weibel~\cite{Weibel.1963} and Horsfield~et~al.~\cite{Parker.1971} using three-dimensional particle tracking-velocimetry (3D PTV). The model, which included up to six bifurcation generations, was used to investigate flow within a range of Reynolds numbers based on the diameter of the trachea, the oscillation frequency and the tidal volume of the airways from $\num{250}$ to $\num{2000}$ and Womersley numbers from $Wo = 1.9$ to $Wo = 5.2$. A sinusoidal breathing cycle was employed using a water-glycerin mixture as the working fluid. By examining the flow field and Lagrangian statistics, i.e., acceleration, torsion, and curvature, three distinct flow regimes based on the Reynolds number could be identified: a four-vortex structure at $Re = \num{250}$, a single pair of counter-rotating vortices for $\num{250} \le Re < \num{1000}$, and two pairs of counter-rotating helices for $Re \ge \num{1500}$. One of the few studies that included the upper airways is the methodological approach by Tauwald et al.~\cite{Tauwald.2024}, which analyzed the flow field inside a patient-specific human nasal cavity model based on a clinical CT scan. The flow field was investigated for physiological breathing cycles using Tomographic Particle-Image Velocimetry (Tomo PIV) with a phased-locked approach and repeated measurements to analyze unsteady flow phenomena during different phases of the breathing cycle. By providing high-spatial resolution flow data in different regions of the nasopharynx, the study emphasizes the relevance of the resulting flow structures induced by the different flow rates and the nasal geometry during the breathing cycle. It was shown that a small spiral-shaped structure affects airflow throughout the entire nasal geometry and surgical procedures in this area could disrupt airflow through the lower nasal turbinates and cause lasting problems.

Although these studies indicate that the geometry of the upper airways, i.e., the nasal and oral cavity significantly influence the flow features in the lower airways, systematic experimental studies quantifying this effect are scarce. Thus, this study focuses on the analysis of the influence of the route of inhalation, i.e. oral or nasal and the flow structures in the lower trachea. To explore this question, the present study employs time-resolved three-dimensional velocity measurements using the Shake-The-Box algorithm in a refractive-index matched patient-specific airway model from the oral and nasal inlets down to the lower trachea, capturing the unsteady oscillatory flow structures that propagate into the primary bifurcation. In a resting state, the tidal volume is about $\SI{500}{\milli\liter}$ per breathing cycle. With a quiet breathing rate of around $\SI{0.1}{\hertz}$, the average airflow is approximately $\SI{300}{\liter\per\hour}$. These conditions result in a tracheal Reynolds number of $Re_{Tr} = 400$. Moderate ventilation, and thus increased tidal volume and breathing frequency, raises the average airway speed, leading to a tracheal Reynolds number of $Re_{Tr} = 1200$. Therefore, steady inhalation with $Re_{Tr} = [400, 1200]$ and oscillatory actuation ($Re_{Tr} = [400, 1200]$ and $Wo = [3, 4.5]$) are used in this study. By keeping the boundary conditions identical for both inhalation routes, the study isolates the impact of oral versus nasal breathing, enabling a direct comparison of velocity distributions and flow structures in the lower trachea.

\section{Experimental Setup and data evaluation}
\label{Sec:expSetup}
The airway geometry used in this study~\cite{Farkas.2020, Lizal.2020} is based on a lung model developed at the Brno University of Technology and was reconstructed from a three-dimensional CT scan of an adult Caucasian male~\cite{Lizal.2012}. This baseline model includes the oral cavity, the laryngeal region, and the tracheobronchial tree down to the 7th generation. To enable a comprehensive assessment of airflow and particle transport during nasal, oral, and combined breathing, the geometry was expanded by incorporating a nasal cavity provided by the University of California, Davis~\cite{Sarangapani.2000}. The nasal model, derived from CT data of a healthy 25-year-old male, was further refined using Rhinoceros 3D~\cite{McNeel.2010} and Star-CCM+, where paranasal sinuses were removed, artefacts corrected, and the surface smoothed to ensure numerical and experimental robustness. Finally, the nasal and oral cavities were anatomically aligned and merged following established morphological references~\cite{putz.2001}. The geometric data was provided by Dr. František Lízal at the Brno University of Technology \citep{Farkas.2020,Lizal.2020}. 

\begin{figure}[H]
     \centering
     \begin{subfigure}[b]{0.32\textwidth}
         \centering
         \includegraphics[height=5cm]{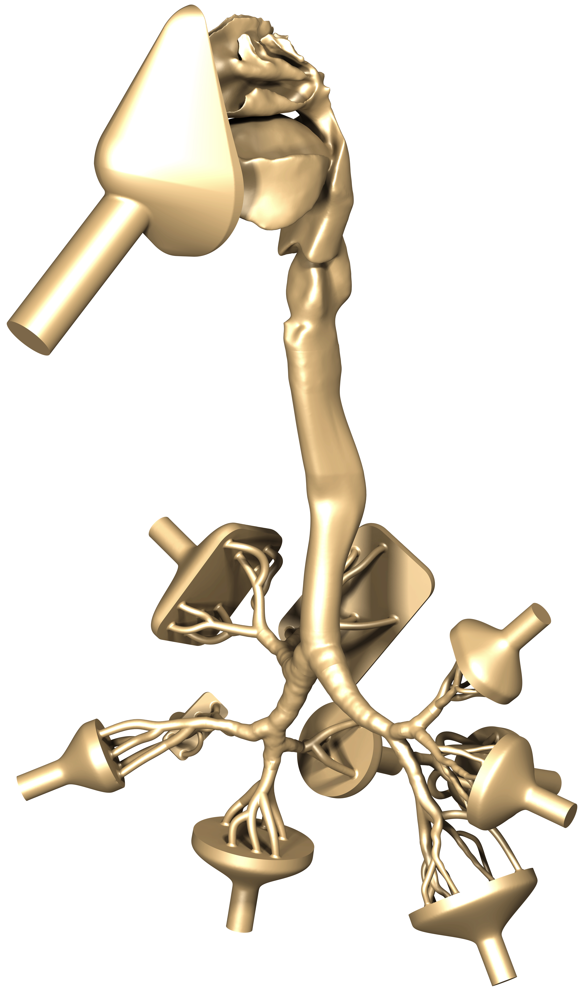}
         \caption{CAD model of the respiratory tract.}
         \label{fig:workflowAirway_CAD}
     \end{subfigure}
     \hfill
     \begin{subfigure}[b]{0.32\textwidth}
         \centering
         \includegraphics[height=5cm]{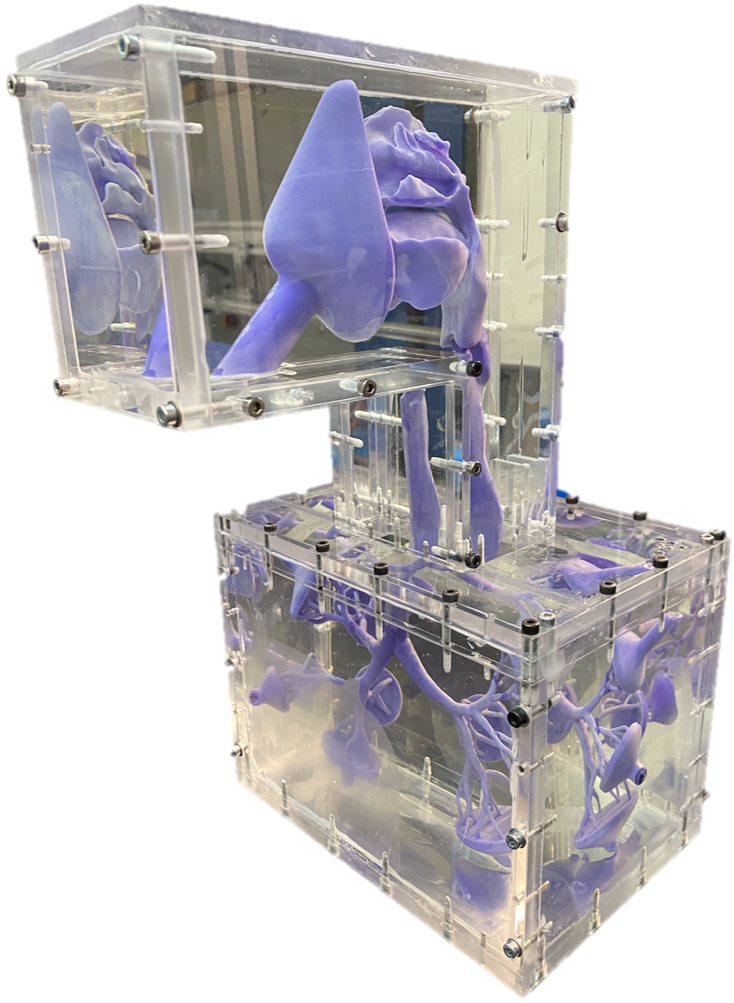}
         \caption{3D printed wax core and cast RTV-615 silicone.}
         \label{fig:workflowAirway_Wax}
     \end{subfigure}
     \hfill
     \begin{subfigure}[b]{0.32\textwidth}
         \centering
         \includegraphics[height = 5cm]{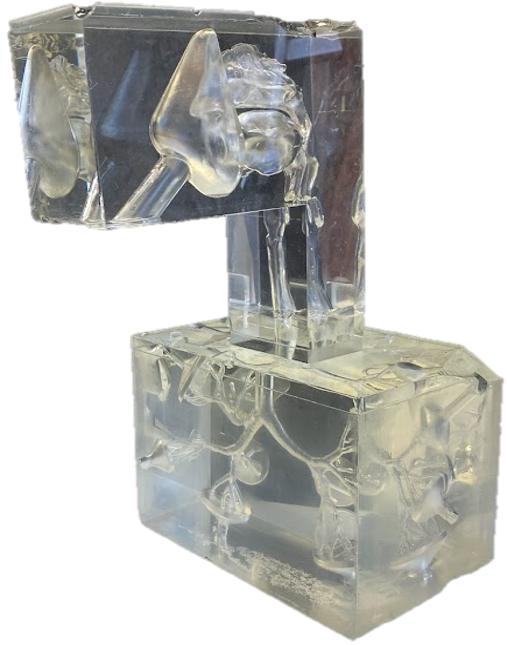}
         \caption{Final silicone model with printing wax removed.}
         \label{fig:workflowAirway_final}
     \end{subfigure}
        \caption{Workflow of the lost-core method for the silicone model of the respiratory tract used during the experiments.}
        \label{fig:workflow_Airways}
\end{figure}

The silicone model was manufactured using the lost-core approach. For this, the model core was printed on an SLA printer (ProJet MJP 2500W) using wax (Visijet M2 Cast) as the printing material, resulting in a spatial resolution of $1200 \times 1200 \times 1600~DPI$ with a layer thickness of \SI{16}{\micro\metre}, which creates a high-precision physical replica of the digital dataset. The wax core of the respiratory model was divided into four parts (oral/nasal, trachea, and left and right lower airways) to reduce the printing time. The lower left and right airways, as well as the oral/nasal cavities and the trachea, were then combined into two parts by soldering. The casting of the silicone model was divided into two stages: 1) the lower airways and 2) the upper airways. The lower airways are placed inside a casting tray which is filled with liquid silicone (RTV-615). After curing the silicone at room temperature for approximately 7~days, the wax core of the trachea is connected to the lower airways, and the final casting tray is assembled and filled with silicone. Finally, the wax core is removed carefully without altering the geometry or mechanical properties of the silicone model. For further information on the manufacturing process, see \cite{JohanningMeiners.2026}. The simplified workflow of the lost-core method is shown in Figure \ref{fig:workflow_Airways}. The mouth and nostrils were covered with a medical face mask during the experiments to provide defined inflow conditions for both inhalation routes. 

\begin{figure}
    \centering
    \includegraphics[width=1\linewidth]{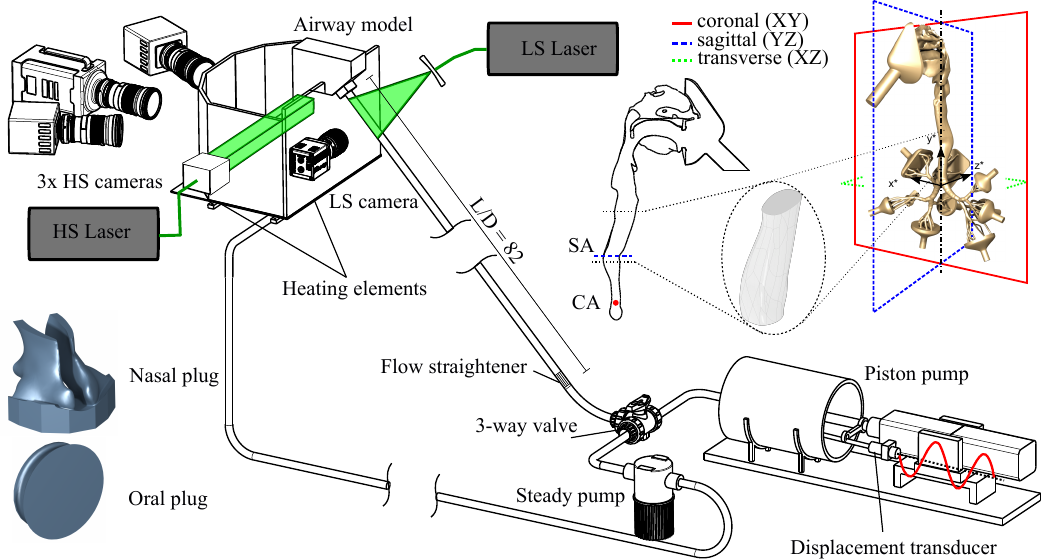}
    \caption{Schematic of the setup for steady inhalation and oscillatory inhalation/exhalation, the coordinate system with corresponding planes and the silicone plugs for individual inhalation.}
    \label{fig:setup}
\end{figure}

Individual inhalation through one cavity (oral or nasal) is achieved by sealing the opposite cavity with a specially designed plug. For this purpose, both seals are manufactured using the lost-core approach. As shown in Figure \ref{fig:setup} bottom left, both plugs precisely follow the original contour of their respective cavity. To minimize the total number of downstream exits within the model, the 52 bronchial tubes of the sixth generation of the bronchial tree are merged into ten funnels. To ensure a quasi-rigid, non-compliant airway replica, the wall thicknesses $\delta_{wall}$ of the upper airways, trachea, and bronchial tree were in the range of $\SI{8}{\milli\meter} < \delta_{wall} < \SI{100}{\milli\meter}$.

A water–glycerin mixture is used to match the refractive index of the silicone model ($n_{Model} = 1.406$), enabling undistorted optical measurements. The water/glycerin mixture consists of \SI{56.75}{\wtpercent} glycerin and \SI{43.25}{\wtpercent} distilled water, resulting in a density of $\rho_{WG} = \SI{1139}{\kg\per\cubic\metre}$ and a dynamic viscosity of $\eta_{WG} = \SI{0.006}{\pascal\second}$ at a temperature of $T_{WG} = \SI{30}{\celsius}$. The temperature of the water/glycerin mixture inside the measurement tank is controlled to ensure constant refractive index, density, and viscosity of the measurement fluid throughout the experiments and is seeded with \textit{Orgasol} particles with a mean diameter of \SI{47.7}{\micro\metre}, resulting in a maximum Stokes number of $St_{max} = \num{5e-4}\ll 1$, based on the seeding particles density and diameter and the viscosity of the fluid. 
Figure \ref{fig:setup} (top right) illustrates the orientation and position of the coordinate system for the PTV and PIV measurements. Normalized coordinates $(x^*,~y^*,~z^*)$ are defined based on the hydraulic diameter of the trachea $d_{Tr} = 4\cdot A_{Tr}/U_{Tr} = \SI{16.3}{\milli\metre}$ with $A_{Tr}$ as the cross-sectional area and $U_{Tr}$ as the circumference of the trachea. The origin of the coordinate system is located at the branching point of the first bifurcation generation with $y^{*}$ pointing upward into the trachea. The $x^{*}$-$y^{*}$ plane corresponds to the coronal plane, $y^{*}$-$z^{*}$ to the sagittal plane, and $x^{*}$-$z^{*}$ to the transverse plane. 

In Figure~\ref{fig:setup}, a schematic of the setup is shown. The airway model is embedded in a closed-circuit flow facility that can generate both steady and oscillatory inhalation and exhalation. Upstream of the medical mask replica, a straight inlet pipe with a hydraulic diameter of $d_i = \SI{20}{\milli\metre}$ and a length of $L_i = \SI{1640}{\milli\metre} = 82 d_i$ is installed to generate defined inflow conditions, i.e., a fully developed velocity profile. Furthermore, a flow straightener is placed directly downstream of the bent pipe and the 3-way valve to reduce the flow disturbances created by the steady pump or the piston pump. The operating mode of the setup can be modified by adjusting the orientation of the three-way valve. 

\begin{figure}[H]
%
%
\definecolor{mycolor1}{rgb}{0.00000,0.44700,0.74100}%
\definecolor{mycolor2}{rgb}{0.85000,0.32500,0.09800}%
\definecolor{mycolor3}{rgb}{0.92900,0.69400,0.12500}%
\definecolor{mycolor4}{rgb}{0.49400,0.18400,0.55600}%
\begin{tikzpicture}

\begin{axis}[%
width=0.4\textwidth,
scale only axis,
xmin=0,
xmax=7,
xtick={0,1.5707963267949,3.14159265358979,4.71238898038469,6.28318530717959},
xticklabels={{0},{$\pi\text{/2}$},{$\pi$},{$\text{3}\pi\text{/2}$},{$\text{2}\pi$}},
xlabel={$\phi$},
ymin=0,
ymax=1.01,
ylabel={normalized piston position},
axis background/.style={fill=white},
xmajorgrids,
ymajorgrids,
legend style={legend cell align=left, align=left, draw=white!15!black}
]
\addplot [color=blue]
  table[row sep=crcr]{%
0.0507527084586397	0.00902451131047319\\
0.152258125375919	0.0232554792189663\\
0.253763542293198	0.0438841491114876\\
0.355268959210477	0.0706336932993401\\
0.456774376127757	0.103145144649433\\
0.558279793045036	0.140982463448387\\
0.659785209962312	0.18363714380016\\
0.761290626879593	0.230537524328787\\
0.862796043796874	0.281054295359867\\
0.964301460714156	0.334508580802256\\
1.06580687763143	0.390184010839149\\
1.16731229454871	0.447333479881974\\
1.26881771146599	0.505189001724758\\
1.37032312838327	0.562975601991958\\
1.47182854530055	0.619916738553915\\
1.57333396221783	0.675248309107238\\
1.67483937913511	0.728228472525596\\
1.77634479605239	0.778145910021493\\
1.87785021296967	0.824330751510263\\
1.97935562988694	0.866163460988859\\
2.08086104680422	0.903082144562699\\
2.1823664637215	0.934591465349196\\
2.28387188063879	0.960268971757176\\
2.38537729755606	0.979769870668209\\
2.48688271447334	0.992832529108831\\
2.58838813139062	0.999281471180714\\
2.6898935483079	0.999030335673891\\
2.79139896522518	0.992082432192525\\
2.89290438214246	0.97853099854942\\
2.99440979905974	0.958557889031522\\
3.09591521597702	0.932430953906168\\
3.1974206328943	0.900501128255363\\
3.29892604981157	0.863196753314015\\
3.40043146672885	0.821018437407656\\
3.50193688364614	0.774532494833264\\
3.60344230056342	0.724362171662713\\
3.70494771748069	0.671181006628983\\
3.80645313439797	0.615702667606668\\
3.90795855131525	0.558671649536252\\
4.00946396823253	0.500853283629021\\
4.11096938514981	0.443023110558833\\
4.21247480206709	0.385957900981373\\
4.31398021898437	0.330423086207138\\
4.41548563590165	0.277163919056389\\
4.51699105281893	0.226895414652877\\
4.6184964697362	0.180291532610469\\
4.72000188665348	0.137977997694622\\
4.82150730357077	0.100522638957419\\
4.92301272048805	0.0684280908606216\\
5.02451813740532	0.042125048141965\\
5.1260235543226	0.0219664860809634\\
5.22752897123988	0.00822285893308105\\
5.32903438815716	0.00107877009312699\\
5.43053980507444	0.000629984932419777\\
5.53204522199172	0.00688252595067684\\
5.633550638909	0.0197524867938974\\
5.73505605582628	0.0390671582407784\\
5.83656147274356	0.0645673458845852\\
5.93806688966083	0.0959108484082605\\
6.03957230657812	0.132676806390904\\
6.1410777234954	0.174372291484242\\
};
\addlegendentry{\small{Reference position}}

\addplot [color=orange, mark=o, mark options={solid, orange}, each nth point=1, filter discard warning=false, unbounded coords=discard,only marks,line width=1pt]
  table[row sep=crcr]{%
0.0507527084586397	0.00735157250561955\\
0.203010833834559	0.033491413697343\\
0.355268959210477	0.0690284859785585\\
0.507527084586397	0.125149921867496\\
0.659785209962312	0.17688080884519\\
0.812043335338237	0.250464359859967\\
0.964301460714156	0.332713476846493\\
1.11655958609007	0.420186109250118\\
1.26881771146599	0.502456728977465\\
1.42107583684191	0.596370757481409\\
1.57333396221783	0.685963110405641\\
1.72559208759375	0.751053610863848\\
1.87785021296967	0.821974417281305\\
2.03010833834559	0.874616178527579\\
2.1823664637215	0.937637387611179\\
2.33462458909742	0.971818566475242\\
2.48688271447334	0.991881597108449\\
2.63914083984926	1.00412328577368\\
2.79139896522518	0.989668980093083\\
2.9436570906011	0.974729574713734\\
3.09591521597702	0.925508345128948\\
3.24817334135294	0.887455936158613\\
3.40043146672885	0.822409279387202\\
3.55268959210477	0.749438303363355\\
3.70494771748069	0.675365747457333\\
3.85720584285661	0.581710733178904\\
4.00946396823253	0.502365091338098\\
4.16172209360845	0.408741602078793\\
4.31398021898437	0.335900463351448\\
4.46623834436028	0.249747466513956\\
4.6184964697362	0.186661383157804\\
4.77075459511213	0.121322157276313\\
4.92301272048805	0.06731158248506\\
5.07527084586397	0.0307674640316881\\
5.22752897123988	0.0145075440407922\\
5.3797870966158	-0.00296285858236407\\
5.53204522199172	0.00692448098046696\\
5.68430334736764	0.0259261635548044\\
5.83656147274356	0.0676443218508325\\
5.98881959811948	0.11250592468923\\
6.1410777234954	0.174351192141283\\
};
\addlegendentry{\small{Piston position}}

\addplot[only marks, mark=*, mark options={}, mark size=3pt, color=black, fill=mycolor3, forget plot] table[row sep=crcr]{%
x	y\\
1.2592	0.5\\
4.00	0.5\\
};
\node[right, align=left, inner sep=0]
at (axis cs:1.4765,0.5) {\small{max inhalation}};
\node[right, align=left, inner sep=0]
at (axis cs:4.24,0.5) {\small{max exhalation}};
\addplot [color=black, forget plot,<->,line width=1pt]
  table[row sep=crcr]{%
1.0526	0.3\\
1.4655	0.3\\
};
\node[right, align=left, inner sep=0]
at (axis cs:1.5708,0.3) {\small{$\Delta\phi=(6/40)\cdot\pi$}};
\addplot [color=white!15!black, forget plot,line width=1pt]
  table[row sep=crcr]{%
1.0527	0\\
1.0527	1.2\\
};
\addplot [color=white!15!black, forget plot,line width=1pt]
  table[row sep=crcr]{%
1.4655	0\\
1.4655	1.2\\
};
\end{axis}
\end{tikzpicture}%
    \caption{Example of the normalized measured piston position and the corresponding prescribed sine. Points of maximum inhalation and exhalation and integration window are indicated.}
    \label{fig:sinewave} 
\end{figure}
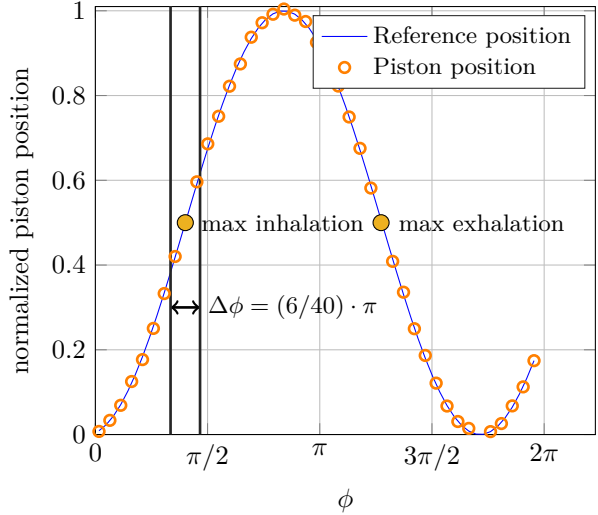

Steady inhalation ($Re_{Tr} = [400,1200]$) is realized by a non-pulsating pump. To determine the volume flux $\dot{V}$ and thereby the Reynolds number in the trachea, the inlet velocity profile upstream of the mask is measured using a low-speed (LS) PIV measurement system that consists of a Continuum Minilite laser and a PCO edge sCMOS camera with a \SI{100}{\milli\metre} lens. The Reynolds number in the trachea is calculated as follows: 
\begin{equation}
Re_{Tr}=\frac{\rho_{W G} \cdot \dot{V}/A_{Tr} \cdot d_{T r}}{\eta_{W G}},
\end{equation} 
where $A_{Tr}$ denotes the area of the trachea. A linear actuator (SEW-EURODRIVE CMS63M coupled with a MOOG G400 series spindle) is used to drive the piston to simulate an oscillating breathing pattern for two Reynolds numbers of $Re_{Tr} = [400,1200]$ and for two Womersley numbers of $Wo = [3, 4.5]$. The oscillatory Reynolds number in the trachea is calculated based on the absolute value of the bulk velocity during peak inhalation. The position of the piston is measured with an accuracy of \SI{2}{\micro\metre} using a magnetostrictive linear sensor that is directly connected to the linear actuator to match the position of the piston pump with the high-speed images. An exemplary measurement of the piston stroke, compared to the pure sine wave used as input, is depicted in Fig.~\ref{fig:sinewave} as a function of the phase angle $\phi$. Note that only every 150th data point is plotted for better clarity. The piston position is normalized with the maximum piston stroke and plotted for a single inhalation and exhalation cycle. The agreement between the specified piston position and the measurement is very good, resulting in a precise and controlled volume flux during inhalation. The point of maximum inhalation is indicated by a yellow circle. An integration window of $\Delta\phi = (6/40)\cdot\pi$ at the points of maximum inhalation/exhalation is chosen to increase the spatial reconstruction resolution. This duration can be considered sufficiently short to assume a steady state, as stated by~Janke~et~al.~\cite{Janke.2017}.
To ensure that the water/glycerin level inside the measurement tank does not change and to keep the geodetic pressure on the outlets of the model at a constant value, the surface area of the tank is designed to be sufficiently large. To minimize additional changes in refractive index and ensure that the lenses of the cameras are positioned parallel to the outer boundaries of the measuring tank, the outer walls of the tank are angled accordingly.

High-speed (HS) images of the flow field in the airways are acquired using three high-speed cameras (one Photron Fastcam NOVA S12 at \SI{0}{\degree} and two Photron Fastcam Mini WX100 at \SI{25}{\degree} orientation relative to the sagittal plane), approximately \SI{800}{\milli\meter} to the measurement volume, equipped with Zeiss \SI{100}{\milli\metre} lenses, with a resolution of $\SI{1024}{\px} \times \SI{1024}{\px}$, yielding a spatial resolution of $\SI{19.5}{\px\per\milli\meter}$. A Quantronix Darwin-Duo~\SI{100}{\milli\joule} Nd:YLF laser with a wavelength of $\lambda = \SI{527}{\nano\metre}$ equipped with a LaVision volume optic is used to illuminate the measurement volume with an approximate size of $20\times55\times30~\si{\cubic\milli\metre}$, which is highlighted in Figure \ref{fig:setup} (top right). For each Reynolds number, a recording frequency of \SI{619}{\hertz} and \SI{1806}{\hertz} was used, respectively. 

A three-dimensional calibration target (LaVision Type~11) was used to establish a mapping between the camera pixels and the real-world coordinates. To ensure reproducible positioning between experimental runs, the airway model was fixed using mechanical brackets. During calibration, the model was removed and the calibration target was placed at the same bracket-defined reference position inside the measurement tank filled with the water/glycerin mixture. The target was positioned at three equally spaced axial locations separated by $\SI{10}{\milli\meter}$. The reference plane was chosen to be located at the center of the tracheal lumen, to ensure consistent placement of the measurement plane within the model across all runs. The mapping procedure was based on refractive-index matching combined with a third-order polynomial fitting model, resulting in a maximum fit error of the volume self-calibration across the three cameras of $\epsilon_{fit} = \SI{0.0099}{\px}$. To further enhance the quality of the raw images, a pre-processing routine was implemented, i.e., subtracting a robust moving average from each image and applying a geometric mask. The pre-processed images of the volumetric measurements are then processed using the software DaVis~10 by LaVision, with a triangulation error of $\epsilon_{tri} = \SI{1.5}{\voxel}$ and a shake width of $\delta_S = \SI{0.11}{\voxel}$. To minimize the influence of ghost particles, tracks with a minimum length smaller than five time steps are not considered. To further minimize the influence of ghost particles, a relatively low seeding density of about $0.03~ppp$ was employed in all experiments given that we could only fit three cameras in the present experiment. Lastly, for Eulerian binning a grid size of \SI{3}{\voxel} or \SI{0.15}{\milli\meter} is applied. The processing parameters are summarized in Table \ref{tab:stb_parameter}.
\begin{table}[H]
    \centering    
    \caption{Shake-The-Box processing parameters.}
    \begin{tabular}{ll}
    \hline
        Allowed triangulation error & $\epsilon_{tri} = \SI{1.5}{\voxel}$\\
        Maximum fit error & $\epsilon_{fit} = \SI{0.0099}{\voxel}$\\
        Shake width  & $\delta_S = \SI{0.11}{\voxel}$\\
        Particle intensity threshold & $T_{int} = 0.1 \cdot I_{avg}$\\
        Minimum track length & 5 time steps\\
        Binning grid size & \SI{3}{\voxel} or \SI{0.15}{\milli\meter}\\
    \end{tabular}
    \label{tab:stb_parameter}
\end{table}

\section{Results}
\label{Sec:Results}
The following section presents the experimental results for inhalation and exhalation in the upper airways, focusing on the influence of the inhalation route (oral/nasal) on the flow field in the trachea. The results are divided into three parts, namely cross-validation of the Tomo-PTV measurement results with PIV measurements, peak inhalation flow field analysis, and maximum exhalation flow characteristics.

\subsection{Cross-validation of the Tomo-PTV results with PIV results}
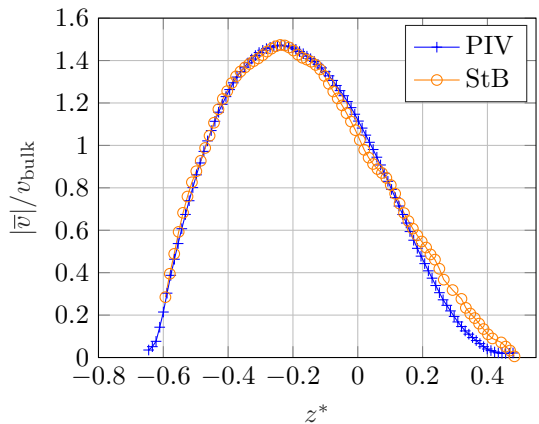
\begin{figure}[H]
%
%
\begin{tikzpicture}

\begin{axis}[%
width=0.35\textwidth,
height=4.5cm,
scale only axis,
xmin=-0.8,
xmax=0.55,
xtick={-0.8, -0.6, -0.4, -0.2, 0,  0.2,    0.4},
xlabel style={font=\color{white!15!black}},
xlabel=$z^*$,
ymin=0,
ymax=1.6,
ytick={  0, 0.2, 0.4, 0.6, 0.8,   1, 1.2, 1.4, 1.6},
ylabel style={font=\color{white!15!black}},
ylabel=$\lvert \overline{v} \rvert/v_{\text{bulk}}$,
axis background/.style={fill=white},
xmajorgrids,
ymajorgrids,
legend style={legend cell align=left, align=left, draw=white!15!black}
]
\addplot [color=blue, mark=+, mark options={solid, blue}]
  table[row sep=crcr]{%
0.479006737	0.0215267135384615\\
0.467651635	0.0200886995384615\\
0.456296504	0.0190020682307692\\
0.444941401	0.0199348146153846\\
0.433586299	0.0220059972307692\\
0.422231197	0.0262824872307692\\
0.410876095	0.0327112546923077\\
0.399520963	0.0402481617692308\\
0.388165861	0.0511281479230769\\
0.376810759	0.0656462489230769\\
0.365455657	0.0800293653846154\\
0.354100555	0.0942000261538462\\
0.342745453	0.110069483076923\\
0.331390321	0.125486073846154\\
0.320035219	0.146146061538462\\
0.308680117	0.167743976153846\\
0.297325015	0.193429824615385\\
0.285969913	0.219805996923077\\
0.274614781	0.246008026923077\\
0.263259679	0.272119560769231\\
0.251904577	0.3057285\\
0.240549475	0.339485963076923\\
0.229194373	0.374132022307692\\
0.217839256	0.408655949230769\\
0.206484154	0.442823395384615\\
0.195129037	0.477971795384615\\
0.183773935	0.514788696153846\\
0.172418833	0.553025476923077\\
0.161063716	0.593835688461538\\
0.149708614	0.634054036923077\\
0.138353512	0.67434523\\
0.126998395	0.714172365384615\\
0.115643293	0.753808365384615\\
0.104288183	0.7924363\\
0.0929330736	0.830559892307692\\
0.0815779716	0.868895761538461\\
0.0702228621	0.907769853846154\\
0.0588677526	0.944299369230769\\
0.0475126468	0.980666853846154\\
0.0361575373	1.01462809230769\\
0.0248024315	1.04838231538462\\
0.0134473238	1.08163219230769\\
0.00209221593	1.11467918461538\\
-0.00926289149	1.14692186153846\\
-0.0206179991	1.17813853076923\\
-0.0319731049	1.20688401538462\\
-0.0433282144	1.23417308461538\\
-0.0546833202	1.25825646153846\\
-0.0660384297	1.28241422307692\\
-0.0773935392	1.3062962\\
-0.0887486413	1.32800203076923\\
-0.100103751	1.34774389230769\\
-0.11145886	1.36525046153846\\
-0.12281397	1.38181299230769\\
-0.134169072	1.39572712307692\\
-0.145524174	1.40857180769231\\
-0.156879291	1.42048712307692\\
-0.168234393	1.43225938461538\\
-0.17958951	1.4426432\\
-0.190944612	1.45299560769231\\
-0.202299714	1.45941583076923\\
-0.213654831	1.46554570769231\\
-0.225009933	1.46925633076923\\
-0.236365035	1.47228665384615\\
-0.247720152	1.47079183846154\\
-0.259075254	1.46852330769231\\
-0.270430356	1.46263344615385\\
-0.281785458	1.45622984615385\\
-0.29314059	1.44845040769231\\
-0.304495692	1.43782977692308\\
-0.315850794	1.42363126153846\\
-0.327205896	1.4072644\\
-0.338560998	1.38924247692308\\
-0.34991613	1.36903077692308\\
-0.361271232	1.34785243846154\\
-0.372626334	1.32226187692308\\
-0.383981436	1.29504123846154\\
-0.395336539	1.26317727692308\\
-0.406691641	1.23007503846154\\
-0.418046772	1.19355458461538\\
-0.429401875	1.15655695384615\\
-0.440756977	1.11359747692308\\
-0.452112079	1.07003461538462\\
-0.463467181	1.02133830769231\\
-0.474822313	0.971762607692308\\
-0.486177415	0.917232323076923\\
-0.497532517	0.861336007692308\\
-0.508887649	0.801131415384615\\
-0.520242751	0.739301396923077\\
-0.531597853	0.674447818461538\\
-0.542952955	0.607291322307692\\
-0.554308057	0.537321544615385\\
-0.565663159	0.463355590769231\\
-0.577018261	0.387440106153846\\
-0.588373363	0.303429106923077\\
-0.599728465	0.214502043076923\\
-0.611083567	0.142919845384615\\
-0.622438669	0.0763125646153846\\
-0.633793831	0.0512747237692308\\
-0.645148933	0.0354134047692308\\
};
\addlegendentry{PIV}

\addplot [color=orange, mark=o, mark options={solid, orange}]
  table[row sep=crcr]{%
0.483462865030675	0.00549173867\\
0.469829588957055	0.0329725250233333\\
0.456196319018405	0.051735034854\\
0.442563049079755	0.07128781592\\
0.415296503067485	0.0899610637\\
0.401663233128834	0.10935716324\\
0.388030018404908	0.1342895731\\
0.374396748466258	0.1599123448\\
0.360763478527607	0.1859303802\\
0.347130202453988	0.2073562802\\
0.333496932515337	0.2341219544\\
0.319863660736196	0.27646815\\
0.292597117791411	0.3180768802\\
0.265330632515337	0.3671171472\\
0.251697360736196	0.4184363768\\
0.238064089570552	0.4616282152\\
0.224430817791411	0.4860278432\\
0.21079754601227	0.5185279848\\
0.197164274233129	0.5470434072\\
0.183531002453988	0.5758530856\\
0.169897760122699	0.6074988126\\
0.156264488343558	0.643387389\\
0.142631217177914	0.68088001\\
0.128997945398773	0.7256795762\\
0.115364703067485	0.7724592804\\
0.101731431288343	0.8131662846\\
0.0880981595092024	0.8467486622\\
0.0744648877300613	0.8699526908\\
0.0608316159509202	0.8884336116\\
0.0471983736196319	0.9124218942\\
0.0335651018404908	0.9422431938\\
0.0199318306748466	0.979137349\\
0.00629855889570546	1.0253503794\\
-0.00733468343558286	1.0720038408\\
-0.0209679552147239	1.112645006\\
-0.0346012269938651	1.151876426\\
-0.0482344987730062	1.188722634\\
-0.0618677558282209	1.221743298\\
-0.075501027607362	1.254184676\\
-0.0891342846625767	1.290352202\\
-0.102767555828221	1.327575326\\
-0.116400812883436	1.358448314\\
-0.130034084662577	1.379308652\\
-0.143667341717791	1.39448211\\
-0.157300613496933	1.404538844\\
-0.17093387791411	1.413088656\\
-0.184567142331288	1.42791009\\
-0.198200406748466	1.44721321\\
-0.211833671165644	1.461754884\\
-0.225466938895706	1.471187168\\
-0.239100203251534	1.473713832\\
-0.252733469447853	1.467484552\\
-0.266366734723926	1.456052964\\
-0.28	1.442250794\\
-0.293633265276074	1.425523044\\
-0.307266530552147	1.41024785\\
-0.320899796748466	1.397352122\\
-0.334533061104294	1.383906126\\
-0.348166328834356	1.367634058\\
-0.361799593251534	1.34861021\\
-0.375432857668712	1.325538254\\
-0.38906612208589	1.293738842\\
-0.402699386503068	1.25595281\\
-0.416332658282209	1.219152068\\
-0.429965915337423	1.170338846\\
-0.443599187116564	1.106777812\\
-0.457232444171779	1.045614649\\
-0.470865715337423	0.9878860488\\
-0.484498972392638	0.9275473008\\
-0.498132244171779	0.878304565\\
-0.511765501226994	0.8261802554\\
-0.525398773006135	0.7590714454\\
-0.539032044785276	0.6822900532\\
-0.552665316564417	0.5922888932\\
-0.566298558895705	0.4887303112\\
-0.579931830674847	0.394941866333333\\
-0.593565101840491	0.284626991\\
};
\addlegendentry{StB}

\end{axis}
\end{tikzpicture}%
    \caption{Comparison of the streamwise velocity profiles in the sagittal plane of the time averaged axial velocity, normalized by the axial bulk velocity $|\overline{v}| / v_{bulk}$ in the trachea at position SA $y^* = 4.2$ for $Re_{Tr} = 400$, extracted from PIV and PTV.}
    \label{fig:validation}
\end{figure}

To validate the Tomo-PTV measurements, velocity profiles inside the trachea were compared with reference PIV data obtained in a previous study \cite{johanningmeiners.2023}. For this purpose, two velocity profiles at the same normalized $y^{*}$ position are compared under steady inhalation conditions at a Reynolds number of $Re_{Tr} = 400$. Figure~\ref{fig:validation} presents the absolute values of the time-averaged axial velocity ($\lvert \overline{v} \rvert$), normalized by the bulk velocity at maximum inspiration ($v_{\text{bulk}}$), for both PIV and PTV measurements in the cross-section SA at $y^{*} = 4.2$. The absolute value of the axial velocity is shown since $y^{*}$ points upward into the trachea. The two measurement techniques showed very good agreement, with only minor deviations around $z^{*} = 0.3$, which likely result from reflections caused by particles adhering to the inner walls of the silicone model. Given the strong agreement between PIV and PTV, all subsequent analyses were based on the volumetric measurements.

\subsection{Flow field during maximum inhalation}
\label{sec:res_inhalation}
\begin{figure}[H]
    \input{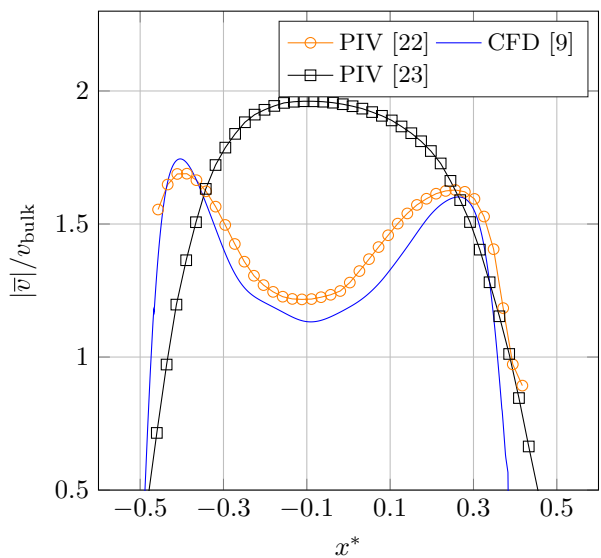}
    \caption{Comparison of the mean axial velocity normalized by the bulk velocity at $y^* = 1.84$ in the trachea of 2D/2C PIV experiments \cite{johanningmeiners.2023} and a CFD simulation \cite{ruettgers.2025} with experiments using a simplified but anatomically shaped inlet trachea \cite{Groe.2007} upstream of the first bifurcation generation.}
    \label{fig:comparison_geometryl}
\end{figure}

To highlight the influence of the upstream conditions, 2D/2C PIV measurements conducted in the same geometry in \cite{johanningmeiners.2023} are compared to the findings of Große~et~al.~\cite{Groe.2007}, who investigated a model starting from the trachea up to the sixth bronchi generation. The position of the CA cross section (indicated by a red dot in Figure~\ref{fig:setup}) is approximately $y^* = y/d_{Tr}=1.84$ above the first bronchi generation, with $d_{Tr}$ as the hydraulic diameter of the trachea. Figure \ref{fig:setup} (top right) indicates the position in the coronal plane. Figure \ref{fig:comparison_geometryl} shows the comparison of the PIV results (dotted red), the CFD data (blue) of R\"uttgers~et~al.~\cite{ruettgers.2025}, and the data of Große~et~al.~\cite{Groe.2007}. The statistically averaged axial velocity $\overline{v}$ is normalized with the mean axial velocity $v_{bulk}$, calculated for the same cross section, and plotted against the normalized axial position $x^* = x/d_{Tr}$. 2Große~et~al.~\cite{Groe.2007} used an anatomically shaped inlet trachea with a length of $l_i = 27 \cdot d_{Tr}$ to generate a fully developed laminar velocity profile as the inlet condition for the trachea. However, as mentioned by Große~et~al.~\cite{Groe.2007}, the flow inside a real human trachea cannot be fully developed due to the short total length of the trachea of only \SIrange{10}{12}{\centi\metre}, i.e., $l_{Tr}/d_{Tr}<10,$, and the three-dimensional structures upstream of the trachea, i.e., the oral/nasal cavity and the larynx and pharynx. As seen in Figure \ref{fig:comparison_geometryl}, the anatomical structures of the airways result in a well-pronounced "m-shaped" profile and not a typical parabolic velocity profile. This direct comparison of the velocity profiles emphasizes the importance of realistic inflow conditions in the first bifurcation and the relevance of the measured volume chosen in this study. 

\begin{figure}[H]
    \centering
    \def\svgwidth{\textwidth}
    \scriptsize
    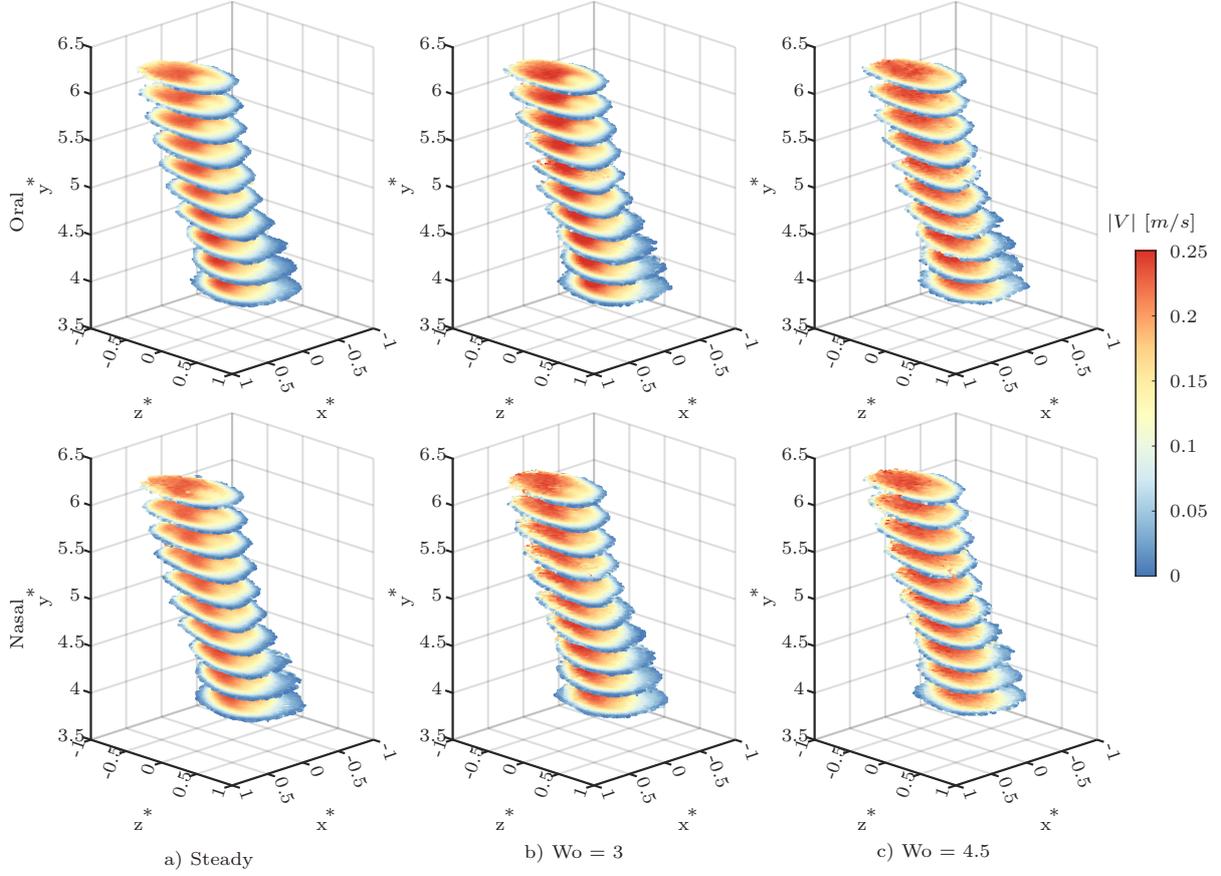   
    \caption{Contour plots of oral (top) and nasal (bottom) inhalation at different $y^*$ locations color coded by the absolute velocity during steady (a) as well as oscillatory flow (b) and (c) for $Re_{Tr} = 400$ and $Wo = [3, 4.5]$.}
    \label{fig:contour_Re400}
\end{figure}

\begin{figure}[H]
    \centering
    \def\svgwidth{\textwidth}
    \scriptsize
    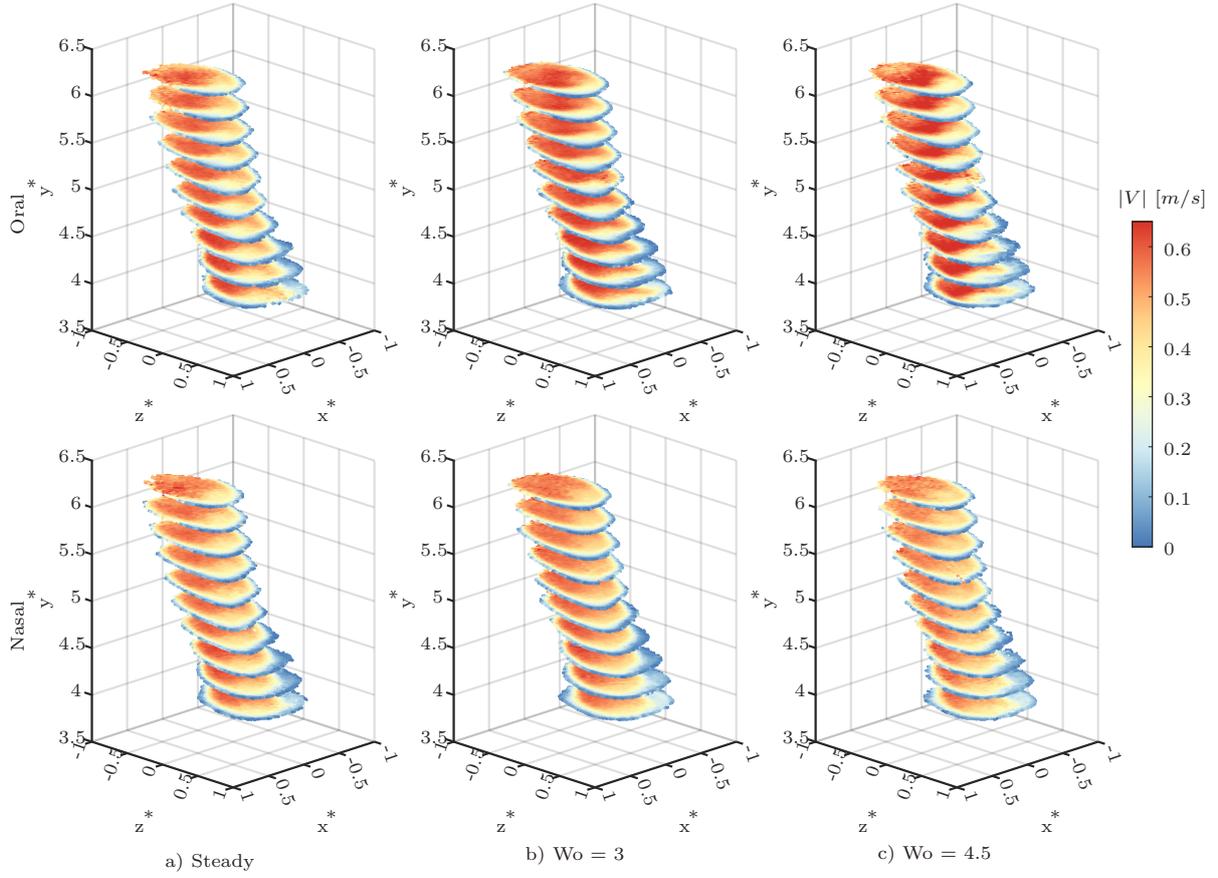   
    \caption{Contour plots of oral (top) and nasal (bottom) inhalation at increasing $y^*$ locations color coded by the absolute velocity during inhalation for $Re_{Tr} = 1200$ and $Wo = [3, 4.5]$.}
    \label{fig:contour_Re1200}
\end{figure}

Figures~\ref{fig:contour_Re400} and~\ref{fig:contour_Re1200} show a comparison of ten equidistant horizontal (transverse) $x^*$-$z^*$ cross sections from $y^*=4$ to $y^*=6.4$ of the absolute velocity magnitude $\lvert V \rvert = \sqrt{u^2+v^2+w^2}$, where $u$, $v$, and $w$ denote the velocity components in the $x^*$-, $y^*$-, and $z^*$-direction, respectively. The contour plots are shown for steady inhalation (left) and peak inhalation for the two Womersley numbers $Wo = 3$ (center) and $Wo = 4.5$ (right) and for oral (top) and nasal (bottom) inhalation. The holes in the velocity contour plots and areas that do not cover the entire measurement volume result from near wall reflections and the settings used for the Shake-The-Box evaluation. Since the allowable triangulation error is limited, not all particle tracks in the rear part of the volume are resolved. 
In Figure~\ref{fig:contour_Re400}, all cross sections for oral and nasal inhalation exhibit a similar velocity contour characterized by a central high-velocity core that diminishes toward the outer walls and forms a distinct arc-shaped pattern downstream. For the oscillatory cases ($Wo = 3$ and $Wo = 4.5$), the overall structure of the velocity field remains comparable to that of the steady case, with minor variations in shape and a broader region of elevated velocity. This broadening is particularly evident at higher Reynolds numbers, where the tracheal jet becomes more diffuse and the shear zones widen due to enhanced momentum exchange and mixing upstream of the first bifurcation. When directly comparing oral and nasal inhalation, no significant differences can be observed, indicating that the spatial development of the flow is largely unaffected by the inhalation route. In Figure~\ref{fig:contour_Re1200}, the corresponding velocity contours for the same cases at an elevated $Re_{Tr} = 1200$ reveal a flow behavior similar to the case of the lower Reynolds number, albeit with a higher magnitude of the absolute velocity. The distinct arc-shaped pattern is less pronounced, and the high-velocity core is broader. Oral inhalation consistently shows slightly higher velocities compared to nasal inhalation, but structural differences between the two inhalation routes are not observed. This result may appear trivial at first glance. However, as seen by the comparison in Figure~\ref{fig:comparison_geometryl}, the velocity profile in the trachea is significantly altered by the presence of the upstream geometries compared to a fully-developed parabolic laminar velocity profile. Consequently, it can be concluded that most likely the pharynx, the larynx, and the redirection of the flow in the throat play a significant role in the spatial development of the flow field in the trachea.

\begin{figure}[H]
    \centering
    \def\svgwidth{\textwidth}
    \scriptsize
    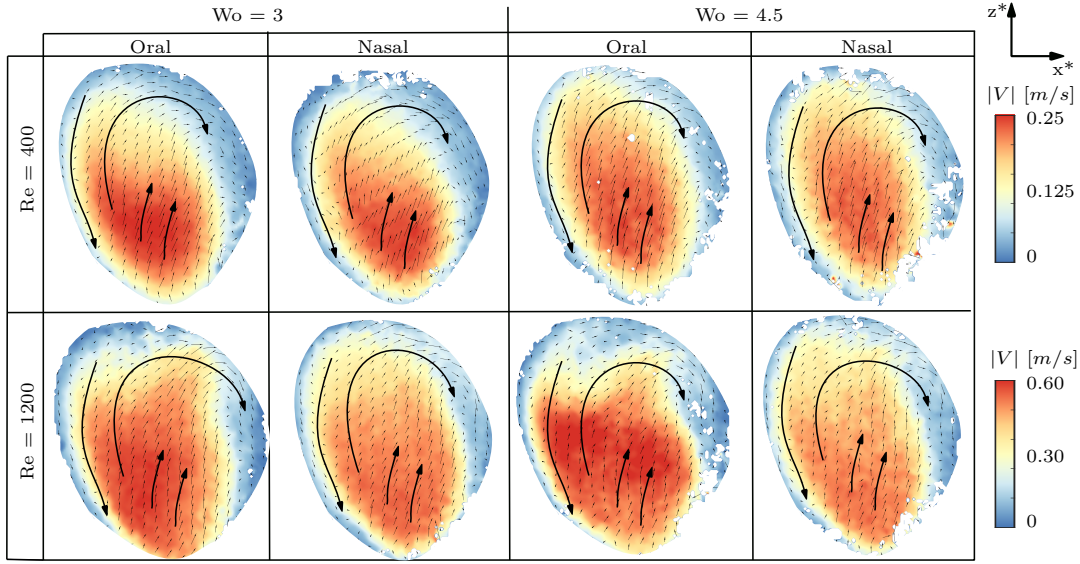   
    \caption{Contour plots in the transverse plane color-coded by the absolute velocity $\lvert V \rvert$ during inhalation for $Re_{Tr} = [400, 1200]$ and $Wo = [3, 4.5]$ at $y^* = 4.2$ for both oral and nasal inhalation.}
    \label{fig:contour_Re400_inhalation}
\end{figure}

\begin{figure}[H]
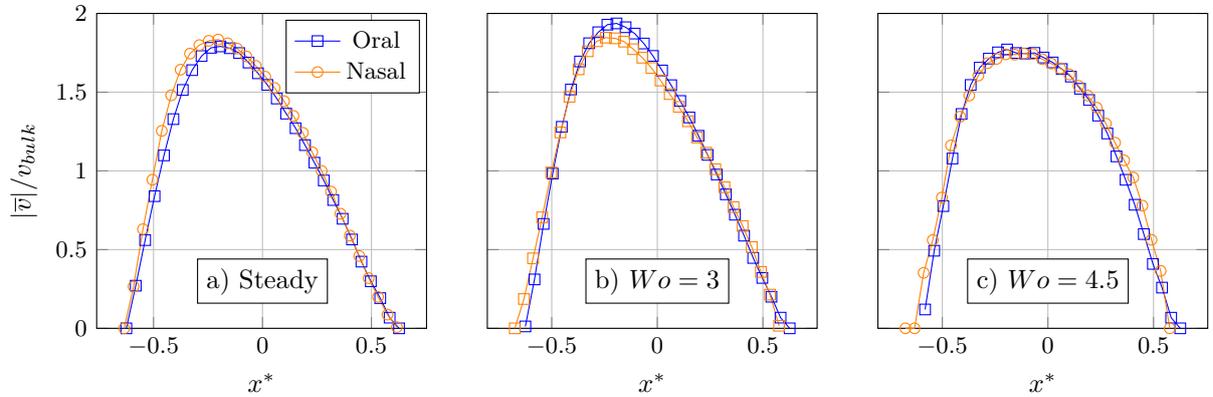

\centering
\begin{tikzpicture}
\begin{groupplot}[
    group style={
        group size=3 by 1,          
        horizontal sep=0.8cm        
    },
    width=0.36\textwidth,            
    height=0.35\textwidth,
    xlabel={$x^*$},                 
    ticklabel style={font=\small},  
    xmajorgrids,
    ymajorgrids,
    ytick={0,0.5,1,1.5,2},   
    ymin=0, ymax=2,          
]
\nextgroupplot[
    ylabel={$|\overline{v}| / v_{bulk}$},  
]
\input{09_Sagittal_Re400_stat}
\nextgroupplot[yticklabel=\empty,
]
\input{09_Sagittal_Re400_Wo3}
\nextgroupplot[  yticklabel=\empty,
    ]
\input{09_Sagittal_Re400_Wo45}
\end{groupplot}
\end{tikzpicture}
\caption{Streamwise velocity profiles in the sagittal plane of the time averaged axial velocity, normalized by the axial bulk velocity $|\overline{v}| / v_{bulk}$ in the trachea at position SA $y^* = 4.2$ for $Re_{Tr} = 400$.}
\label{fig:Re400_sagittal}
\end{figure}

The contour plots of the absolute velocity in a single sagittal plane at $y^{*} = 4.2$ are shown in Figure~\ref{fig:contour_Re400_inhalation}. Similarly to the results of Figures~\ref{fig:contour_Re400} and \ref{fig:contour_Re1200}, the overall flow characteristics of the contour plot and vector field remain consistent for oral and nasal inhalation while the differences in shape and magnitude occur mainly at the higher Reynolds number. The center of the high-speed tracheal jet is located at lower values of $z^*$ on the front side of the trachea for $Re_{Tr} = 400$, while the area becomes slightly larger and is shifted towards higher $z^*$ in the case of a higher Reynolds number due to an enhanced momentum exchange. The influence of the Womersley number is small compared to the influence of the tracheal shape. 
Lastly, a direct comparison is shown in Figure~\ref{fig:Re400_sagittal} in the form of streamwise velocity profiles normalized by the axial bulk velocity $|\overline{v}|/v_{bulk}$ in the sagittal and coronal plane at $y^{*} = 4.2$ for oral and nasal inhalation. The plots show the velocity profiles for steady and oscillatory flow during maximum inhalation from left to right for Womersley numbers of $Wo = [3, 4.5]$. 
For a Reynolds number of $Re_{Tr} = 400$, the three profiles resemble a parabolic shape for nasal and oral inhalation which are slightly skewed towards negative $x^*$. Due to the inclination of the trachea, the velocity profiles are similar to an angled pipe flow, and thus again emphasize the relevance of the physiological geometry of the upper airways. The maximum value nearly approaches the theoretical value of $|\overline{v}| / v_{bulk} = 2$, which is expected for an ideal Hagen-Poiseuille flow in a pipe. For shorter periods, as in the case of $Wo = 4.5$, the profile becomes slightly wider indicating greater sensitivity to the Womersley number at lower Reynolds numbers. 

\begin{figure}[H]
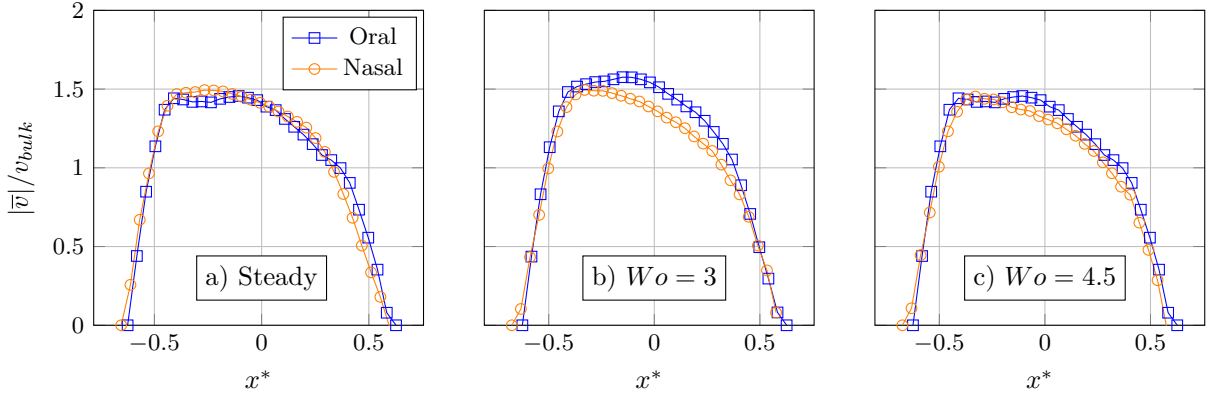

\centering
\begin{tikzpicture}
\begin{groupplot}[
    group style={
        group size=3 by 1,          
        horizontal sep=0.8cm        
    },
    width=0.36\textwidth,            
    height=0.35\textwidth,
    xlabel={$x^*$},                 
    ticklabel style={font=\small},  
    xmajorgrids,
    ymajorgrids,
    ytick={0,0.5,1,1.5,2},   
    ymin=0, ymax=2,
]
\nextgroupplot[
    ylabel={$|\overline{v}| / v_{bulk}$},  
]
\input{10_Sagittal_Re1200_stat}
\nextgroupplot[yticklabel=\empty,
]
\input{10_Sagittal_Re1200_Wo3}
\nextgroupplot[  yticklabel=\empty,
    ]
\input{10_Sagittal_Re1200_Wo45}
\end{groupplot}
\end{tikzpicture}
\caption{Velocity profiles in the sagittal plane of the time averaged axial velocity, normalized by the axial bulk velocity in the trachea $v_{bulk}$ at position SA $y^* = 4.2$ for $Re_{Tr} = 1200$.}
\label{fig:Re1200_sagittal}
\end{figure}

For the higher Reynolds number of $Re=1200$ in Figure \ref{fig:Re1200_sagittal}, the three velocity profiles are more flat and the influence of the Womersley number becomes less prominent. An increase in the Reynolds number enhances momentum exchange and mixing, which results in the jet becoming more diffuse with broader shear layers. Consequently, the flow's sensitivity to the Womersley number is reduced. The profiles possess a steeper gradient resulting in a fuller profile. The plots again show that there appears to be little difference between steady inhalation and oscillatory flow during peak inhalation. For the two Reynolds numbers, the difference between oral and nasal inhalation seems to be negligible. 

\begin{figure}[H]
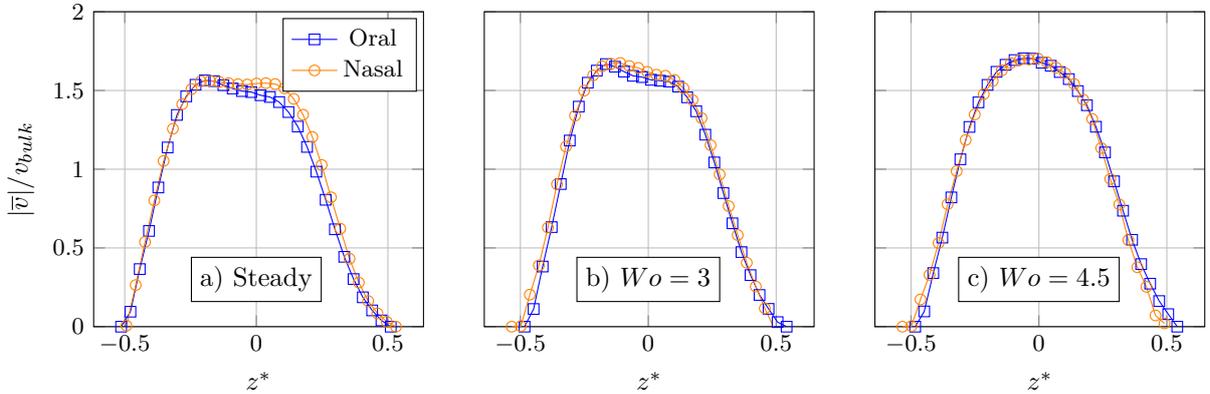

\centering
\begin{tikzpicture}
\begin{groupplot}[
    group style={
        group size=3 by 1,          
        horizontal sep=0.8cm        
    },
    width=0.36\textwidth,            
    height=0.35\textwidth,
    xlabel={$z^*$},                 
    ticklabel style={font=\small},  
    xmajorgrids,
    ymajorgrids,
    ytick={0,0.5,1,1.5,2},   
    ymin=0, ymax=2,
]
\nextgroupplot[
    ylabel={$|\overline{v}| / v_{bulk}$},  
]
\input{11_Coronal_Re400_stat}
\nextgroupplot[yticklabel=\empty,
]
\input{11_Coronal_Re400_Wo3}
\nextgroupplot[  yticklabel=\empty,
    ]
\input{11_Coronal_Re400_Wo45}
\end{groupplot}
\end{tikzpicture}
\caption{Velocity profiles in the coronal plane of the time averaged axial velocity, normalized by the axial bulk velocity in the trachea $v_{bulk}$ at position SA $y^* = 4.2$ for $Re_{Tr} = 400$.}
\label{fig:Re400_coronal}
\end{figure}

\begin{figure}[H]
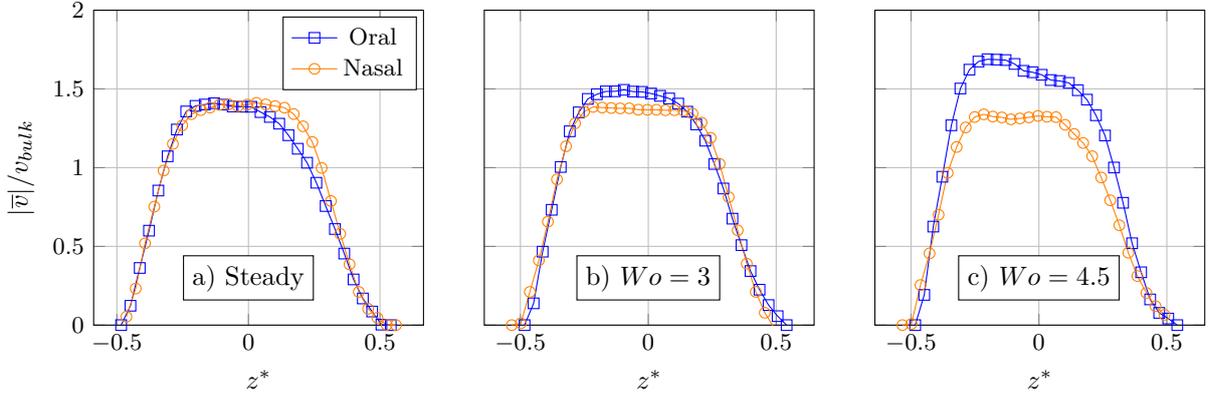

\centering
\begin{tikzpicture}
\begin{groupplot}[
    group style={
        group size=3 by 1,          
        horizontal sep=0.8cm        
    },
    width=0.36\textwidth,            
    height=0.35\textwidth,
    xlabel={$z^*$},                 
    ticklabel style={font=\small},  
    xmajorgrids,
    ymajorgrids,
    ytick={0,0.5,1,1.5,2},   
    ymin=0, ymax=2,
]
\nextgroupplot[
    ylabel={$|\overline{v}| / v_{bulk}$},  
]
\input{12_Coronal_Re1200_stat}
\nextgroupplot[yticklabel=\empty,
]
\input{12_Coronal_Re1200_Wo3}
\nextgroupplot[  yticklabel=\empty,
    ]
\input{12_Coronal_Re1200_Wo45}
\end{groupplot}
\end{tikzpicture}
\caption{Velocity profiles in the coronal plane of the time averaged axial velocity, normalized by the axial bulk velocity in the trachea $v_{bulk}$ at position SA $y^* = 4.2$ for $Re_{Tr} = 1200$.}
\label{fig:Re1200_coronal}
\end{figure}

Figures \ref{fig:Re400_coronal} and \ref{fig:Re1200_coronal} show the corresponding velocity profiles in the coronal plane for the same parameter set. For $Re=400$ (see Figure~\ref{fig:Re400_coronal}) the profiles for oral and nasal inhalation are again very similar, although a stronger influence of the Womersley number can be observed for both cases. While the maxima of the profiles in the steady case and for $Wo = 3$ are slightly shifted towards negative values of $z^{*}$, the velocity profile for $Wo = 4.5$ is more parabolic and centered around $z^*=0$. The difference between oral and nasal inhalation is negligible across all three cases.

The effect of a higher Reynolds number is less pronounced in comparison to the sagittal plane. However, as shown in Figure~\ref{fig:Re1200_coronal}, the asymmetry of the curve diminishes, resulting in a more symmetric profile, while the gradient near the wall stays similar to cases with lower Reynolds numbers. 

For oral inhalation at $Re_{Tr} = 1200\text{ and }Wo=4.5$, the velocity profile shows an increased magnitude near the center compared to nasal inhalation. In Figure~\ref{fig:Re1200_coronal}, for $Wo=4.5$, a difference of approximately \SI{20}{\percent} of the maximum magnitude is observed at $x^*\approx-0.2$. When comparing the results to the contour plot in Figure~\ref{fig:contour_Re1200}, it becomes evident that the the maximum amplitude for oral inhalation is higher, while the core of lower velocities is much broader for nasal inhalation. By comparing the effective volume flux in the $x^*$-$z^*$ plane at $y^*=4.2$, a difference of less than \SI{6}{\percent} can be calculated. Consequently, for higher Reynolds number and Womersley number, the velocity profiles flatten more strongly and the velocity core becomes wider.
In summary, as emphasized by the asymmetric profiles observed in both measurement planes, the influence of the geometry should not be neglected, as a strong influence on the velocity profiles resulting from the shape of the upper airways and the trachea is observed. However, the difference between the oral and nasal inhalation routes in all scenarios is rather small and therefore does not have significant relevance for the downstream flow development in the lower airways, i.e., the bronchial tree.

\subsection{Flow field during maximum exhalation}
\label{sec:res_exhalation}
\begin{figure}[H]
    \centering
    \def\svgwidth{\textwidth}
    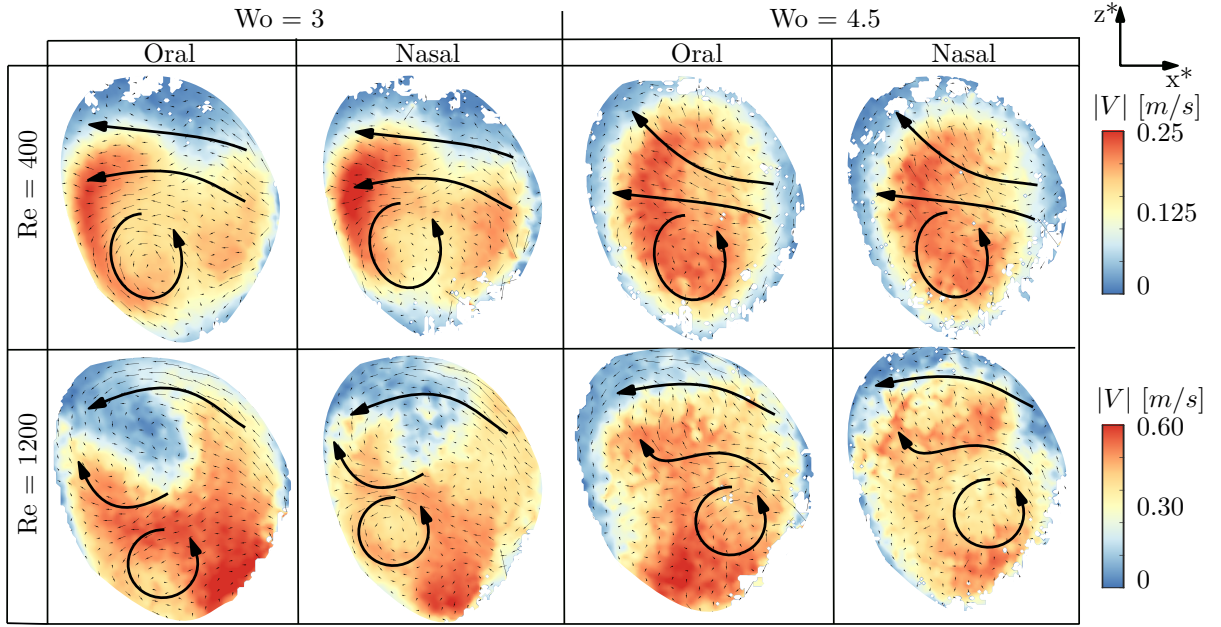   
    \caption{Contour plots in the transverse plane color coded by the absolute velocity during exhalation for $Re_{Tr} = [400, 1200]$ and $Wo = [3, 4.5]$ at $y^* = 4.2$.}
    \label{fig:contour_Re400_exhalation}
\end{figure}

\begin{figure}[H]
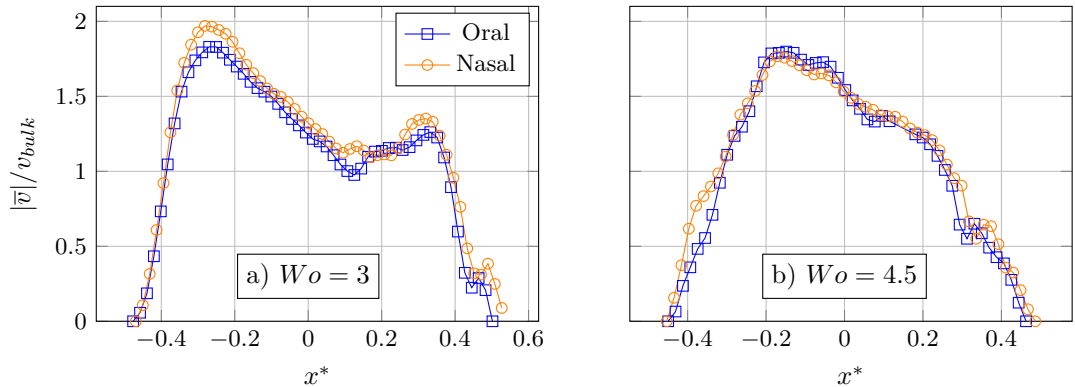

\centering
\begin{tikzpicture}
\begin{groupplot}[
    group style={
        group size=2 by 1, 
        horizontal sep=1.2cm
    },
    width=0.45\textwidth, 
    height=0.35\textwidth,
    xlabel={$x^*$},
    ticklabel style={font=\small},
    xmajorgrids,
    ymajorgrids,
    ytick={0,0.5,1,1.5,2},   
    ymin=0,  
    ymax=2.1,
    each nth point=2,
]

\nextgroupplot[
    ylabel={$|\overline{v}| / v_{bulk}$},
]
\input{14_Re400_Wo3_exhalation}

\nextgroupplot[
    yticklabel=\empty,
]
\input{14_Re400_Wo45_exhalation}
\end{groupplot}
\end{tikzpicture}

\caption{Velocity profiles in the coronal plane of the time averaged axial velocity, normalized by the axial bulk velocity in the trachea $v_{bulk}$ at position SA $y^* = 4.2$ for $Re_{Tr} = 400$.}
\label{fig:Re400_coronal_exhalation}
\end{figure}

While the analysis of the velocity profiles during inhalation already showed that the route of breathing does not introduce noticeable differences in the tracheal flow field, an even smaller influence is expected during exhalation. Since the geometric influence is negligible during exhalation, as both oral and nasal flows originate from the bronchial tree, the tracheal flow fields are expected to be similar. Figure~\ref{fig:contour_Re400_exhalation} shows the velocity contours at $y^*=4.2$ in the transverse plane during peak exhalation. Again, the difference between oral and nasal exhalation routes is minimal for all cases investigated. Since the flow originates from the lower airways, i.e., the bronchial tree, it is not affected by geometrical differences in the upper airways. However, the impact of the Womersley number becomes more pronounced during exhalation. Although the vortex orientation stays consistent for all the cases, the location and magnitude of the high-speed jet as well as the positions of the vortices vary for a higher Reynolds and Womersley number. The high-speed region predominantly moves towards negative $x^*$, suggesting that the majority of the volume flux originates from the left main bronchus. Although the vector field and contour plots during inhalation appear orderly and consistent, the opposite holds true for exhalation. The vector field is characterized by the existence of vortices and a wide range of secondary flow structures. 
Figure~\ref{fig:Re400_coronal_exhalation} shows the velocity profiles in the coronal plane during peak exhalation at $Re_{Tr}=400$. The velocity profiles for oral and nasal airflow are again nearly identical for both Womersley numbers. Both cases show slightly asymmetric profiles around the central axis $(x^*=0)$. The differences in the overall shape of the velocity profiles arise from the change in Womersley number. While the profile has a double-peak or asymmetric "m-shape" for $Wo=3$, the profile at $Wo=4.5$ shows a single but skewed and shifted peak. It can be summarized that the difference in the inhalation/exhalation route during the maximum volume flux is negligible. The difference in the shape of the velocity profiles with increasing Womersley number can be attributed to a phase difference between parallel flow paths. Similar to findings in the literature, e.g., \cite{Jalal.2020}, 
the difference between the two Womersley numbers can be attributed to unequal inertial responses of the lower bifurcation branches due to differences in their length and geometry. Consequently, local differences in inertial impedance can lead to a redistribution of velocities and the observed transition from the double-peaked to the single-skewed profile.

\section{Conclusions}
\label{Sec:conclusion}
This study examined the impact of oral and nasal breathing routes, Reynolds and Womersley numbers, and realistic upper airway geometries on tracheal flow. A patient-specific silicone model of the airways was embedded in a closed-loop setup using a water–glycerin mixture to match its refractive index, enabling distortion-free measurements. High-resolution Tomo-PTV using the Shake-The-Box algorithm fully resolved the flow field in the trachea during oral and nasal inhalation and exhalation. Experiments were conducted under steady and oscillatory inhalation/exhalation, simulating physiologically relevant breathing patterns with Reynolds numbers of $Re_{Tr} = 400$ and $1200$ and Womersley numbers of $Wo=3~\text{and}~4.5$.

The experimental results show that the upstream anatomy strongly shapes the inflow, producing asymmetric, three-dimensional velocity profiles that persist into the trachea and first bifurcation. Compared to idealized, fully developed inflow conditions, anatomically realistic inflow generates asymmetric and m-shaped velocity profiles in the coronal plane. During inhalation, a central high-velocity core forms along the tracheal axis, surrounded by slower near-wall regions, creating an arc-shaped profile in coronal planes. This structure is similar for both oral and nasal routes, indicating minimal impact of the inhalation pathway on gross flow patterns. Direct comparison of the velocity profiles of steady and oscillating flow for oral and nasal inhalation show only minor differences in the tracheal flow field. Yet, it was found that the Reynolds number is the primary factor influencing the flow behavior, with higher Reynolds numbers yielding flatter velocity profiles and more pronounced high-velocity regions. Finally, the Womersley number also affects the shape of oscillatory velocity profiles, though its influence is smaller. During exhalation, the flow is largely independent of the inhalation route, as the flow originates from the lower airways. Nevertheless, variations in Reynolds and Womersley numbers still affect the vortex distribution and the magnitude of the high-speed regions. 

In summary, realistic inlet conditions are essential for capturing physiologically relevant airflow, but the choice of breathing route has a weaker influence on tracheal flow development. These insights are directly applicable to respiratory flow modeling, with implications for predicting particle deposition, optimizing inhalation therapies and medical devices, and developing efficient yet physiologically accurate, computational models.


\section*{Acknowledgements}
The authors gratefully acknowledge the German Research Foundation (Deutsche Forschungsgemeinschaft DFG) for funding this work (project number: 449867589). We also gratefully acknowledge the contribution and support of Trutz Meyer, Nick Capellmann, and the workshop team of the Institute of Aerodynamics during the manufacturing process. The authors are very grateful for the measurement system provided by the Chair and Institute of Aerospace Systems at RWTH Aachen University. The authors are grateful to Dr. Franti\v{s}ek L\'{i}zal and his team at Brno University of Technology for providing the three-dimensional digital model geometry of the respiratory tract.

\section*{Declarations}
\subsection*{Funding}
The research was made possible because of the funding from the Deutsche Forschungsgemeinschaft DFG (German Research Foundation) from the project: "Velocity and wall-shear stress measurements of pulsatile flow of blood-analog fluids in elastic vessels", project number: 449867589. DFG reference number \mbox{KL 2138/8-1}.

\subsection*{Conflict of interest}
The authors declare that they have no known competing financial interests or personal relationships that could have appeared to influence the work reported in this paper.

\subsection*{Consent for publication}
All authors agreed with the content. The authors and the institute authorities gave explicit consent to submit this article.  

\subsection*{Author contribution}
B.JM. and M.K. contributed to conceptualization of the work. B.JM. and L.M. contributed to implementation, experiments, methodology, visualization and writing (original draft and review/editing). M.K. contributed to methodology, supervision, resource and project management and writing (review/editing). D.K. provided advise from a general fluid mechanics perspective. All authors read and approved the final version of the manuscript.

 \bibliographystyle{elsarticle-num} 
 \bibliography{Bib}

@article{McNeel.2010,
  title={Rhinoceros 3D, Version 6.0},
  author={McNeel, Robert and others},
  journal={Robert McNeel \& Associates, Seattle, WA},
  year={2010}
}

@article{Lizal.2012,
author = {Lizal, Frantisek and Elcner, Jakub and Hopke, Philip and Jedelsky, Jan and Jicha, Miroslav},
year = {2012},
month = {03},
pages = {197-207},
title = {Development of a realistic human airway model},
volume = {226},
journal = {Proceedings of the Institution of Mechanical Engineers. Part H, Journal of Engineering in Medicine},
doi = {10.1177/0954411911430188}
}

@book{putz.2001,
  title={Atlas of Human Anatomy Sobotta},
  author={Putz, Reinhard and Pabst, Reinhard and Putz, Renate and Weiglein, Andreas H},
  year={2001},
  publisher={Lippincott Williams \& Wilkins}
}

@article{Sarangapani.2000,
 author = {Sarangapani, Ramesh},
 year = {2000},
 title = {{Modeling Particle Deposition in Extrathoracic Airways}},
 pages = {72--89},
 volume = {32},
 number = {1},
 issn = {0278-6826},
 journal = {{Aerosol Science \& Technology}},
 doi = {10.1080/027868200303948}
}

@misc{ruettgers.2025,
 author = {R{\"u}ttgers, Mario and Vorspohl, Julian and Mayolle, Luca and Johanning-Meiners, Benedikt and Krug, Dominik and Klaas, Michael and Meinke, Matthias and Lee, Sangseung and Schr{\"o}der, Wolfgang and Lintermann, Andreas},
 year = {2025},
 title = {{Comparative Analysis of the Flow in a Realistic Human Airway}},
 publisher = {arXiv},
 doi = {10.48550/arXiv.2510.14320}
}

@article{JohanningMeiners.2026,
 author = {Johanning-Meiners, Benedikt Harald and Klaas, Michael},
 year = {2026},
 title = {{Scalable high-precision silicone models for refractive-index-matched measurements in biomedical applications}},
 volume = {67},
 number = {2},
 issn = {0723-4864},
 journal = {{Experiments in Fluids}},
 doi = {10.1007/s00348-025-04168-w}
}

@article{Parker.1971,
 author = {Parker, H. and Horsfield, K. and Cumming, G.},
 year = {1971},
 title = {Morphology of distal airways in the human lung},
 pages = {386--391},
 volume = {31},
 number = {3},
 issn = {0021-8987},
 journal = {Journal of applied physiology},
 doi = {10.1152/jappl.1971.31.3.386}
}

@book{Weibel.1963,
 author = {Weibel, Ewald R.},
 year = {1963},
 title = {Morphometry of the Human Lung},
 address = {Berlin, Heidelberg},
 publisher = {{Springer Berlin Heidelberg}},
 isbn = {9783642875533},
 doi = {10.1007/978-3-642-87553-3}
}

@article{Wang.2021,
author = {Chia C. Wang  and Kimberly A. Prather  and Josué Sznitman  and Jose L. Jimenez  and Seema S. Lakdawala  and Zeynep Tufekci  and Linsey C. Marr },
title = {Airborne transmission of respiratory viruses},
journal = {Science},
volume = {373},
number = {6558},
pages = {eabd9149},
year = {2021},
doi = {10.1126/science.abd9149},
}

@article{Farkas.2020,
 author = {Farkas, {\'A}rp{\'a}d and Lizal, Frantisek and Jedelsky, Jan and Elcner, Jakub and Karas, Jakub and Belka, Miloslav and Misik, Ondrej and Jicha, Miroslav},
 year = {2020},
 title = {{The role of the combined use of experimental and computational methods in revealing the differences between the micron-size particle deposition patterns in healthy and asthmatic subjects}},
 pages = {105582},
 volume = {147},
 issn = {00218502},
 journal = {Journal of Aerosol Science},
 doi = {10.1016/j.   jaerosci.2020.105582}
}

@article{Lizal.2020,
  author = {Lizal, Frantisek and Elcner, Jakub and Jedelsky, Jan and Maly, Milan and Jicha, Miroslav and Farkas, {\'A}rp{\'a}d and Belka, Miloslav and Rehak, Zdenek and Adam, Jan and Brinek, Adam and Laznovsky, Jakub and Zikmund, Tomas and Kaiser, Jozef},
 year = {2020},
 title = {{The effect of oral and nasal breathing on the deposition of inhaled particles in upper and tracheobronchial airways}},
 pages = {105649},
 volume = {150},
 issn = {00218502},
 journal = {Journal of Aerosol Science},
 doi = {10.1016/j.   jaerosci.2020.105649 }
}

@article{Xu.2021,
 author = {Xu, Jiaquan and Murphy, Sherry L. and Kochanek, Kenneth D. and Bastian, Brigham and Arias, Elizabeth},
 year = {2021},
 title = {{Deaths: Final Data for 2019}},
 journal = {National Center for Health Statistics}
}

@article{Kulkarni.2016,
 author = {Kulkarni, Hemant and Smith, Claire Mary and Lee, Dani Do Hyang and Hirst, Robert Anthony and Easton, Andrew J. and O'Callaghan, Chris},
 year = {2016},
 title = {{Evidence of Respiratory Syncytial Virus Spread by Aerosol. Time to Revisit Infection Control Strategies?}},
 pages = {308--316},
 volume = {194},
 number = {3},
 journal = {American Journal of Respiratory and Critical Care Medicine},
 doi = {10.1164/rccm.201509-1833OC  }
}

@article{Milton.2012,
 author = {Milton, Donald K.},
 year = {2012},
 title = {{What was the primary mode of smallpox transmission? Implications for biodefense}},
 pages = {150},
 volume = {2},
 journal = {Frontiers in Cellular and Infection Microbiology},
 doi = {10.3389/fcimb.2012.00150    }
}

@article{Tang.2005,
 author = {Tang, J. W. and Eames, I. and Li, Y. and Taha, Y. A. and Wilson, P. and Bellingan, G. and Ward, K. N. and Breuer, J.},
 year = {2005},
 title = {{Door-opening motion can potentially lead to a transient breakdown in negative-pressure isolation conditions: the importance of vorticity and buoyancy airflows}},
 pages = {283--286},
 volume = {61},
 number = {4},
 issn = {0195-6701},
 journal = {The Journal of Hospital Infection},
 doi = {10.1016/j.            jhin.2005.05.017 }
}

@article{Groe.2007,
 author = {Gro{\ss}e, S. and Schr{\"o}der, W. and Klaas, M. and Kl{\"o}ckner, A. and Roggenkamp, J.},
 year = {2007},
 title = {{Time resolved analysis of steady and oscillating flow in the upper human airways}},
 pages = {955--970},
 volume = {42},
 number = {6},
 issn = {0723-4864},
 journal = {Experiments in Fluids},
 doi = {10.1007/s00348-007-0318-y            }
}

@article{Xu.2020,
 author = {Xu, Xiaoyu and Wu, Jialin and Weng, Wenguo and Fu, Ming},
 year = {2020},
 title = {{Investigation of inhalation and exhalation flow pattern in a realistic human upper airway model by PIV experiments and CFD simulations}},
 pages = {1679--1695},
 volume = {19},
 number = {5},
 journal = {Biomechanics and Modeling in Mechanobiology},
 doi = {10.1007/s10237-020-01299-3        }
}

@article{Banko.2015            ,
 author = {Banko, A. J. and Coletti, F. and Schiavazzi, D. and Elkins, C. J. and Eaton, J. K.},
 year = {2015},
 title = {Three-dimensional inspiratory flow in the upper and central human airways},
 volume = {56},
 number = {6},
 issn = {0723-4864},
 journal = {Experiments in Fluids},
 doi = {10.1007/s00348-015-1966-y}
}

@article{Janke.2017,
 author = {Janke, Thomas and Schwarze, R{\"u}diger and Bauer, Katrin},
 year = {2017},
 title = {Measuring three-dimensional flow structures in the conductive airways using 3D-PTV},
 volume = {58},
 number = {10},
 issn = {0723-4864},
 journal = {Experiments in Fluids},
 doi = {10.1007/s00348-017-2407-x }
}

@article{Jalal.2020,
 author = {Jalal, Sahar and {van de Moortele}, Tristan and Amili, Omid and Coletti, Filippo},
 year = {2020},
 title = {Steady and oscillatory flow in the human bronchial tree},
 volume = {5},
 number = {6},
 journal = {Physical Review Fluids},
 doi = {10.1103/PhysRevFluids.5.063101 }
}

@article{Koullapis.2018,
 author = {Koullapis, P. and Kassinos, S. C. and Muela, J. and Perez-Segarra, C. and Rigola, J. and Lehmkuhl, O. and Cui, Y. and Sommerfeld, M. and Elcner, J. and Jicha, M. and Saveljic, I. and Filipovic, N. and Lizal, F. and Nicolaou, L.},
 year = {2018},
 title = {Regional aerosol deposition in the human airways: The SimInhale benchmark case and a critical assessment of in silico methods},
 pages = {77--94},
 volume = {113},
 journal = {European Journal of Pharmaceutical Sciences : Official Journal of the European Federation for Pharmaceutical Sciences},
 doi = {10.1016/j.ejps.2017.09.003  }
}

@article{Soodt.2012,
 author = {Soodt, Thomas and Schr{\"o}der, Franka and Klaas, Michael and {van Overbr{\"u}ggen}, Timo and Schr{\"o}der, Wolfgang},
 year = {2012},
 title = {Experimental investigation of the transitional bronchial velocity distribution using stereo scanning PIV},
 pages = {709--718},
 volume = {52},
 number = {3},
 issn = {0723-4864},
 journal = {Experiments in Fluids},
 doi = {10.1007/s00348-011-1103-5  }
}

@article{johanningmeiners.2023,
  title={Combined Velocity and Aerosol Deposition Measurements in the Respiratory Tract using {HS-PIV} and {MTT} Assay},
   year = {2023},
author={Johanning-Meiners, BH and Schier, C and Meyer, T and D{\"o}rner, P and Gruhlke, MCH and Rink, L and Schr{\"o}der, W and Klaas, M},
journal = {15th International Symposium on Particle Image Velocimetry}
}

@article{Tauwald.2024,
 author = {Tauwald, Sandra Melina and Erzinger, Florian and Quadrio, Maurizio and R{\"u}tten, Markus and Stemmer, Christian and Krenkel, Lars},
 year = {2024},
 title = {Tomo-PIV in a patient-specific model of human nasal cavities: a methodological approach},
 pages = {055203},
 volume = {35},
 number = {5},
 issn = {0957-0233},
 journal = {Measurement Science and Technology},
 doi = {10.1088/1361-6501/ad282c }
}

\end{document}